# Gamma Resonances near Threshold for the Production of Thermal Photoneutrons

Silviu Olariu and Agata Olariu

*National Institute of Physics and Nuclear Engineering, Magurele, Romania*

Abstract

We have determined the positions of the (γ,n) resonances and upper limits for the integrated cross sections for the (γ,n) reactions, using data for the inverse process (n,γ). With the aid of these data we have estimated the number of low-energy neutrons which can be generated by the irradiation of a target with a γ-ray beam. Among the reactions producing thermal neutrons via (γ,n) reaction we mention $^{185}$Re(γ,n)$^{184}$Re with an upper limit of the integrated cross section of 2.4 b-eV, and $^{178}$Hf(γ,n)$^{177}$Hf with an upper limit of the integrated cross section of 0.9 b-eV.

## 1. Introduction

In this work we consider an experiment in which a γ-ray beam induces nuclear electromagnetic transitions in a target, in the vicinity of the threshold $S_n$ for the reaction (γ,n).

We determine the positions of the (γ,n) resonances and upper limits for the integrated cross sections for the (γ,n) reactions, using data for the inverse process (n,γ). We list the (γ,n) resonances and the upper limit for the (γ,n) reaction cross sections, in the vicinity of the reaction threshold. With the aid of these data we shall estimate the number of low-energy neutrons which can be generated by the irradiation of a target as a function of the spectral intensity of the γ-ray beam.

## 2. Determination of (γ,n) resonances from the inverse (n,γ) process

The positions of the gamma-neutron resonances (γ,n) above threshold have been determined using existing data for the inverse process (n,γ), with the aid of the relations between the variables characterizing the direct and inverse processes.

The positions of the (γ,n) resonances and upper limits to the cross section of the (γ,n) reaction have been determined for 73 stable isotopes and 101 unstable isotopes, for the energy intervals where we could find data for the (n,γ) reactions.

---

contact: olariu@nipne.ro, agata@nipne.ro

We have used these results in order to identify the cases for which a (γ,n) resonance has a position such that the neutron emitted backwards with respect to the direction of incidence of the gamma-ray photon may have zero energy [1].

Among the reactions for which there are such resonances we mention $^{185}$Re(γ,n)$^{184}$Re with an upper limit of the integrated cross section of 2.4 b-eV, and $^{178}$Hf(γ,n)$^{177}$Hf with an upper limit of the integrated cross section of 0.9 b-eV, in all directions.

## 2.1 Stable isotopes

The stable nuclei for which we could find thermal-neutron resonances are listed in Table I up to energies of the emitted neutrons of 0.1 eV. Upper limits for the corresponding integrated cross sections, with the emission of neutrons into 4π, are also given in Table I.

| Reaction | Upper limit for the integrated cross section, b-eV | Position of resonance $E_\gamma - S_n$, eV | Width of resonance, eV | Height of resonance, b | Lower limit of the neutron energy $D_-$, eV | Upper limit of the neutron energy $D_+$, eV |
|---|---|---|---|---|---|---|
| $^{153}$Eu(γ,n)$^{152}$Eu | 0.007 | 258.2 | 0.187 | 4.973E-2 | 1.588E-4 | 6.704 |
| $^{170}$Yb(γ,n)$^{169}$Yb | 0.036 | 227.9 | 0.145 | 1.666E-1 | 3.348E-4 | 5.295 |
| $^{185}$Re(γ,n)$^{184}$Re | 2.438 | 171.5 | 0.151 | 1.081E1 | 9.013E-4 | 3.837 |
| $^{152}$Sm(γ,n)$^{151}$Sm | 0.121 | 242.6 | 0.124 | 7.559E-1 | 9.863E-4 | 6.557 |
| $^{178}$Hf(γ,n)$^{177}$Hf | 0.971 | 176.5 | 0.088 | 7.657E0 | 0.00197 | 4.156 |
| $^{153}$Eu(γ,n)$^{152}$Eu | 0.044 | 258.4 | 0.224 | 1.462E-1 | 0.00396 | 7.101 |
| $^{156}$Gd(γ,n)$^{155}$Gd | 0.069 | 252.8 | 0.150 | 3.176E-1 | 0.01791 | 7.190 |
| $^{153}$Eu(γ,n)$^{152}$Eu | 0.010 | 257.9 | 0.281 | 3.756E-2 | 0.02136 | 6.030 |
| $^{152}$Sm(γ,n)$^{151}$Sm | 0.212 | 242.9 | 0.116 | 1.299E0 | 0.02259 | 7.181 |
| $^{154}$Sm(γ,n)$^{153}$Sm | 1.735 | 223.2 | 0.106 | 1.188E1 | 0.02957 | 6.659 |
| $^{193}$Ir(γ,n)$^{192}$Ir | 0.059 | 168.5 | 0.106 | 3.786E-1 | 0.04058 | 2.798 |
| $^{164}$Dy(γ,n)$^{163}$Dy( | 0.653 | 193.7 | 0.126 | 3.501E0 | 0.04557 | 5.699 |
| $^{152}$Sm(γ,n)$^{151}$Sm | 0.270 | 242.0 | 0.102 | 1.811E0 | 0.05238 | 5.292 |
| $^{139}$La(γ,n)$^{138}$La | 0.091 | 300.7 | 0.142 | 4.502E-1 | 0.06052 | 10.155 |
| $^{153}$Eu(γ,n)$^{152}$Eu | 0.014 | 257.4 | 0.142 | 6.860E-2 | 0.06719 | 5.488 |
| $^{170}$Yb(γ,n)$^{169}$Yb | 0.210 | 227.4 | 0.098 | 1.307E0 | 0.07073 | 4.216 |
| $^{193}$Ir(γ,n)$^{192}$Ir | 0.102 | 169.5 | 0.114 | 6.265E-1 | 0.09640 | 4.773 |
| $^{170}$Yb(γ,n)$^{169}$Yb | 0.227 | 228.7 | 0.109 | 1.204E0 | 0.09671 | 6.919 |
| $^{150}$Sm(γ,n)$^{149}$Sm | 0.417 | 229.2 | 0.082 | 3.358E0 | 0.09758 | 4.694 |

Table I. Stable-isotope (γ,n) resonances and upper limits for the integrated cross sections for the (γ,n) reaction, with the emission of the neutrons into 4π. $D_-$ represents the energy of the neutron emitted backwards with respect to the direction of incidence, and $D_+$ the energy of the neutron emitted along the direction of incidence of the γ-ray photon.

The (γ,n) resonances for several representative nuclei are shown in Figs. 1-9.

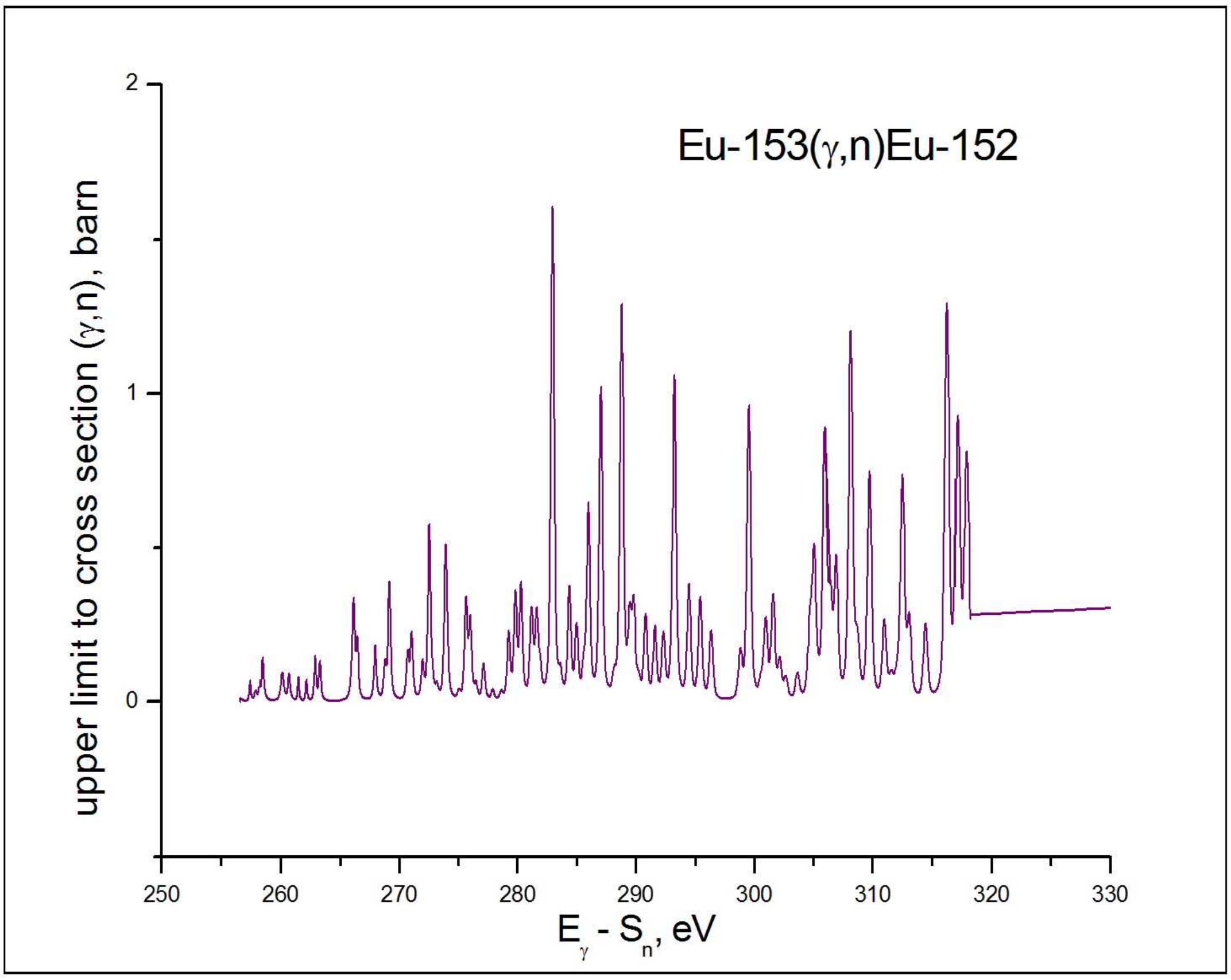

Fig. 1. (γ,n) resonances and upper limits to the integrated cross section for the reaction $^{153}Eu(\gamma,n)\,^{152}Eu$.

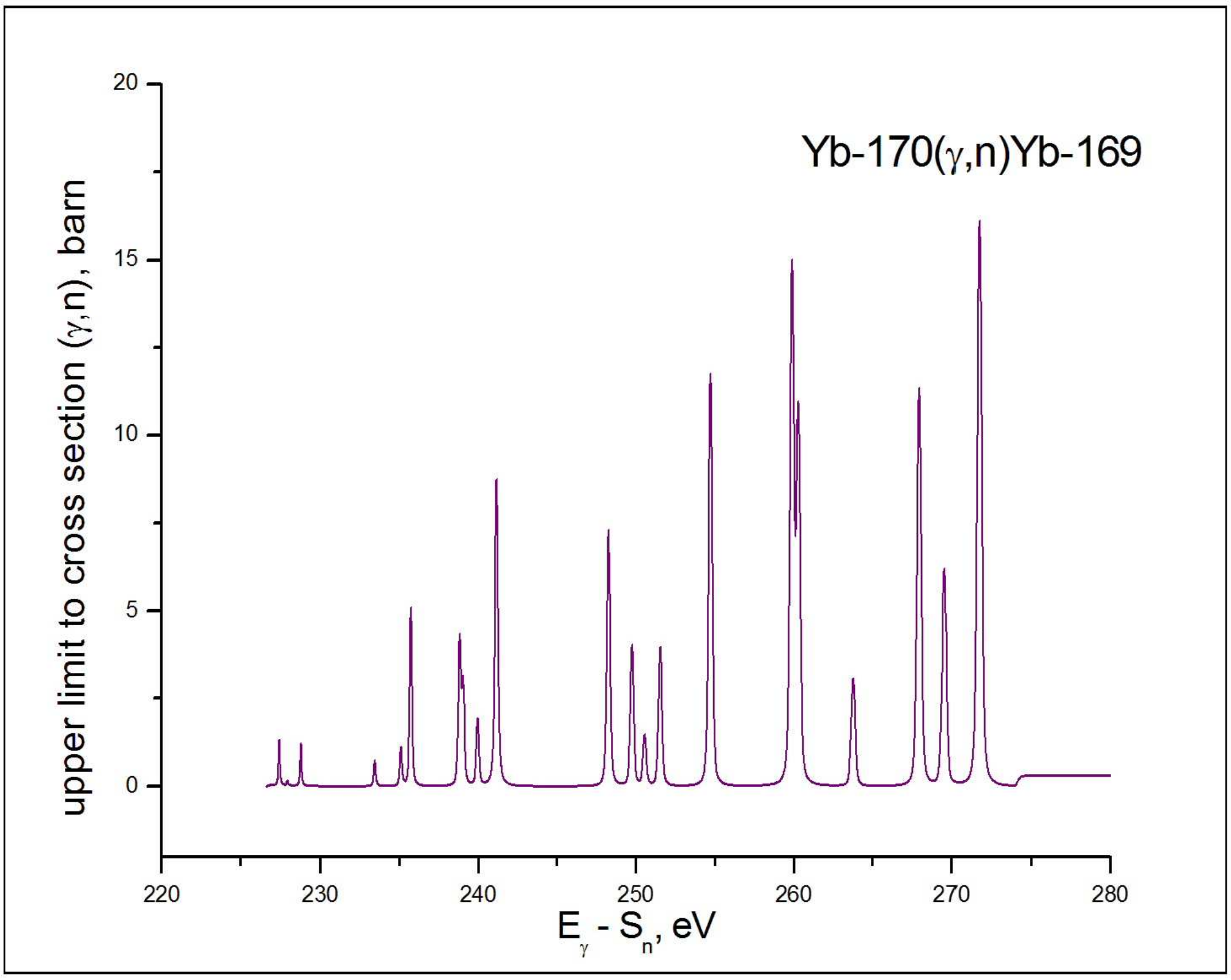

Fig. 2. (γ,n) resonances and upper limits to the integrated cross section for the reaction $^{170}Yb(\gamma,n)^{169}Yb$.

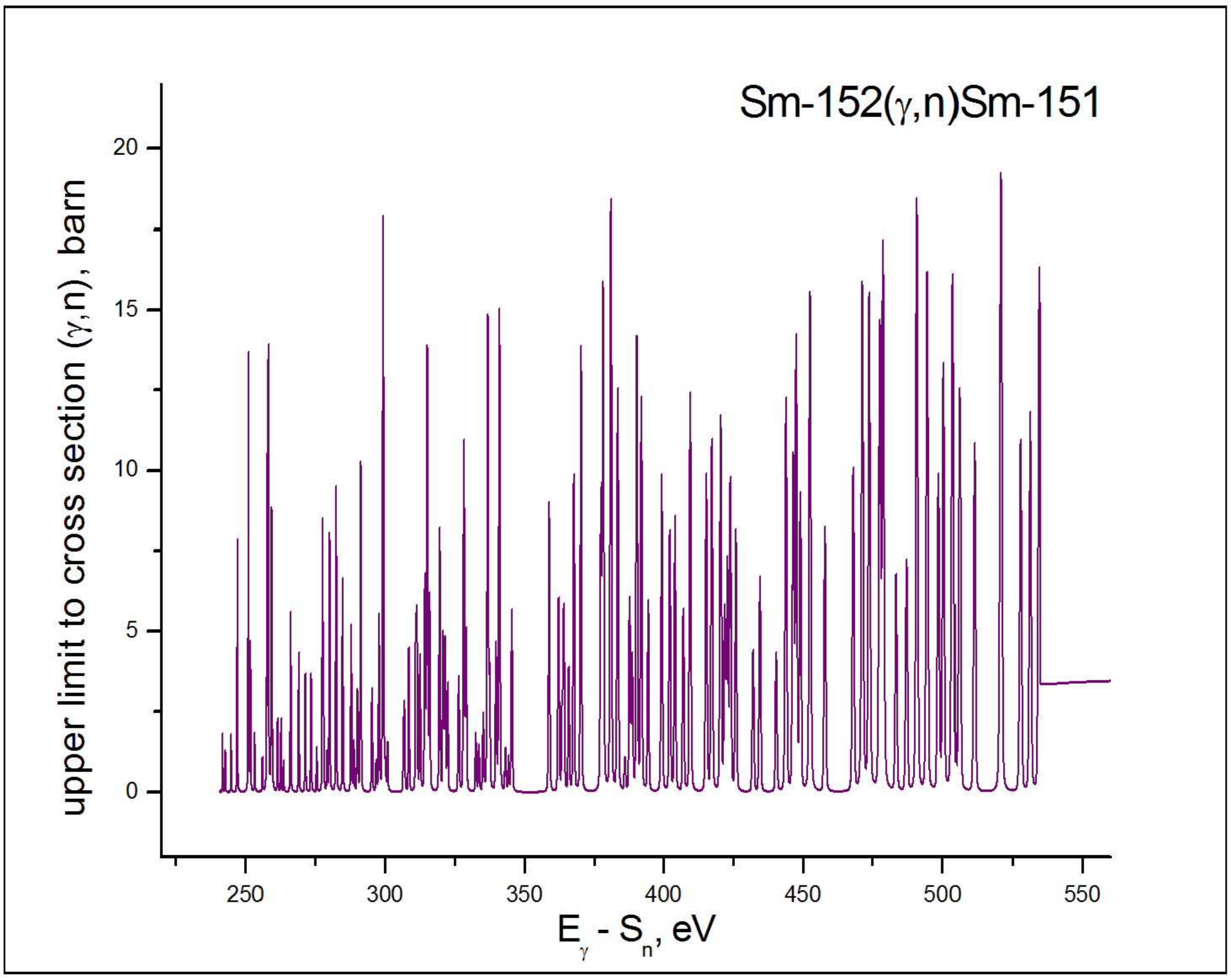

Fig. 3. (γ,n) resonances and upper limits to the integrated cross section for the reaction $^{152}Sm(\gamma,n)\,^{151}Sm$.

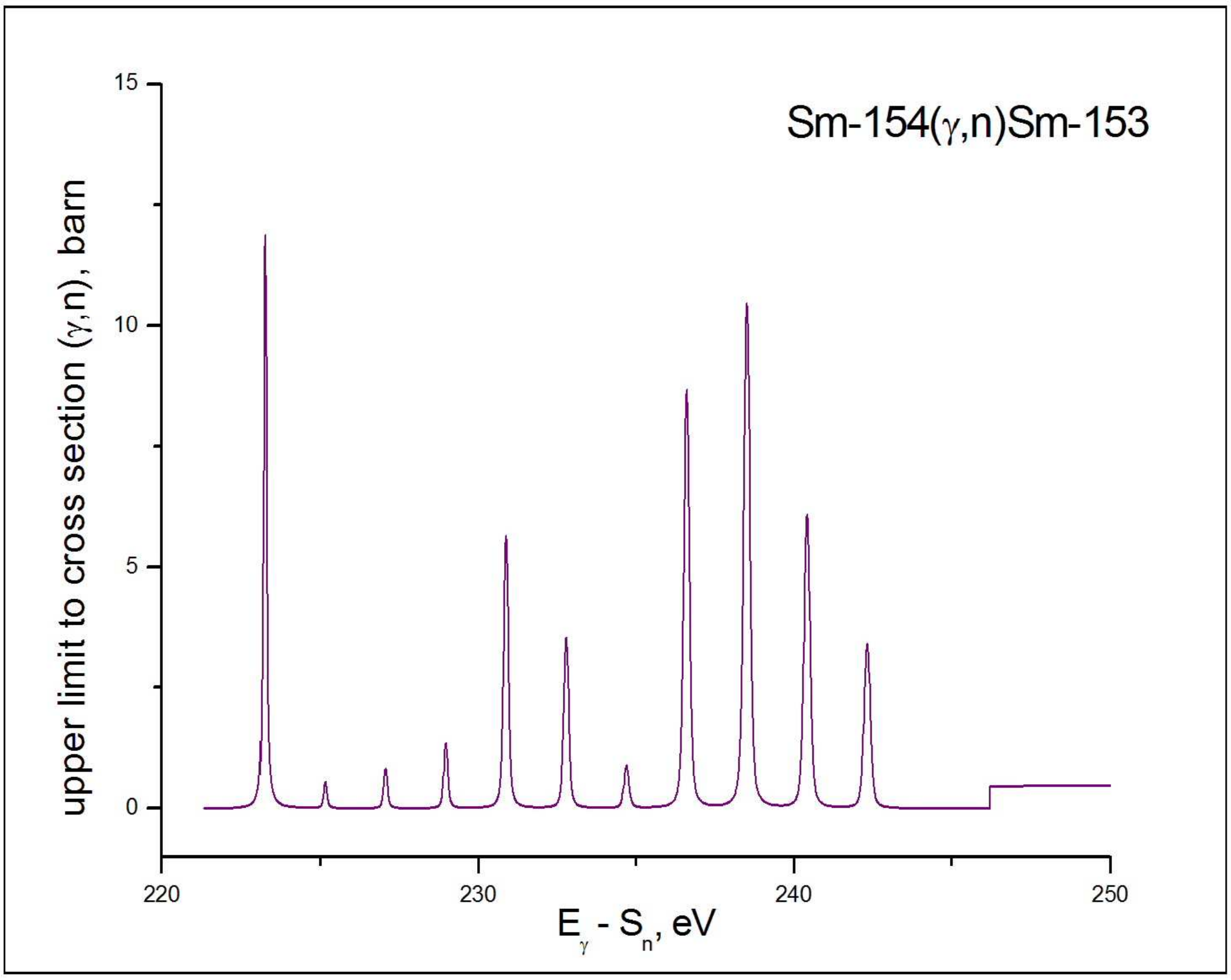

Fig. 4. (γ,n) resonances and upper limits to the integrated cross section for the reaction $^{154}Sm(\gamma,n)\,^{153}Sm$.

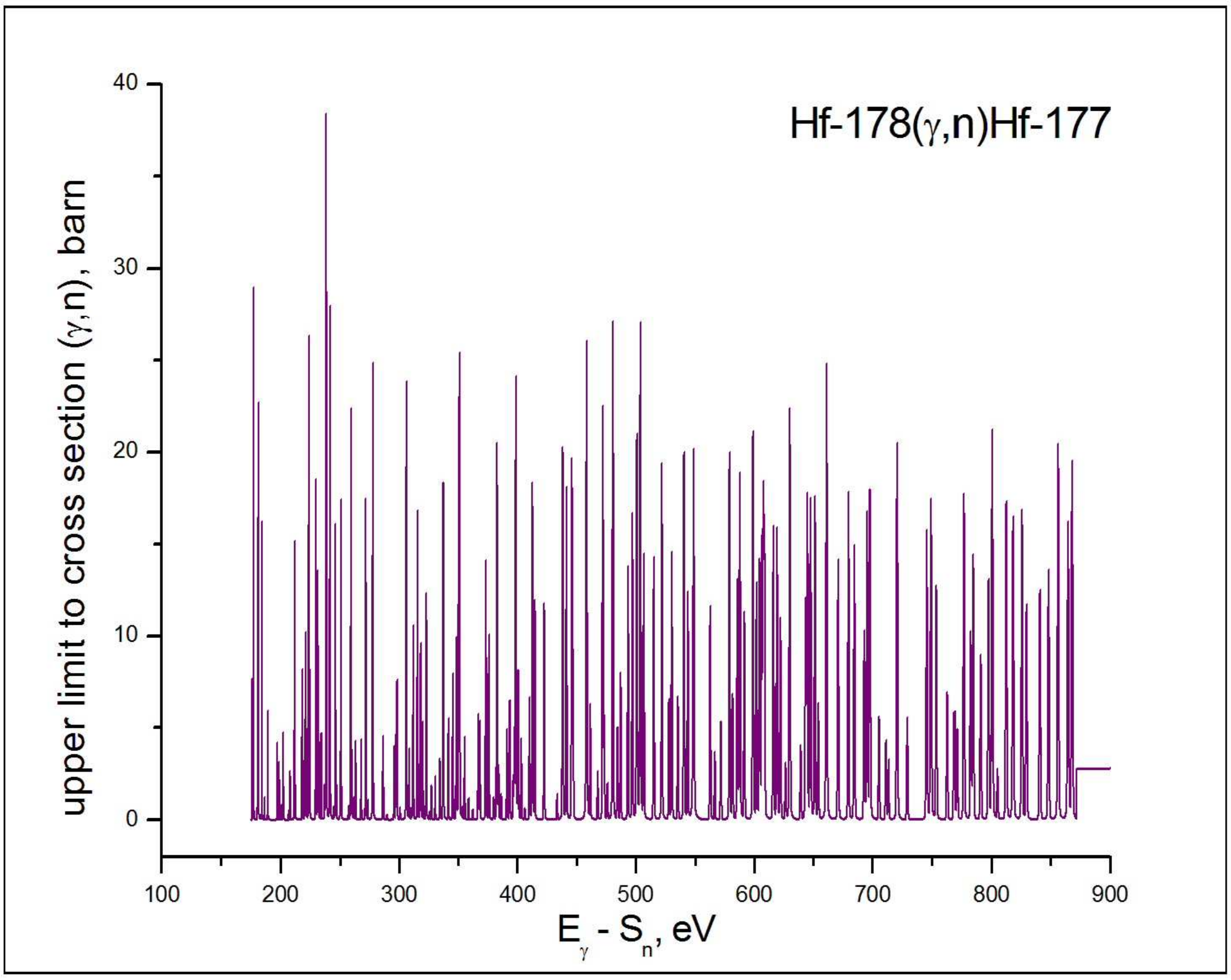

Fig. 5. (γ,n) resonances and upper limits to the integrated cross section for the reaction $^{178}Hf(\gamma,n)^{177}Hf$.

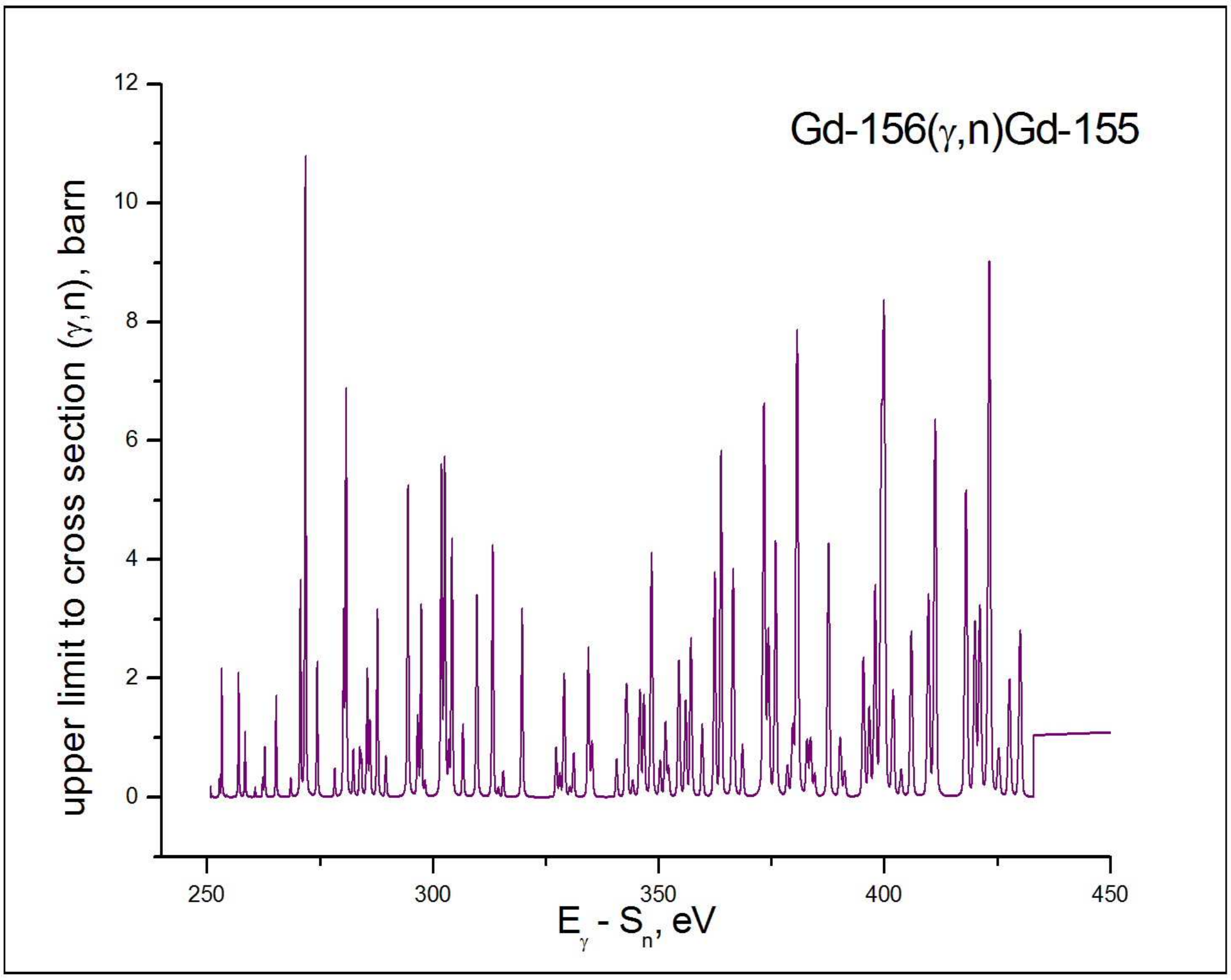

Fig. 6. (γ,n) resonances and upper limits to the integrated cross section for the reaction $^{156}Gd(\gamma,n)^{155}Gd$.

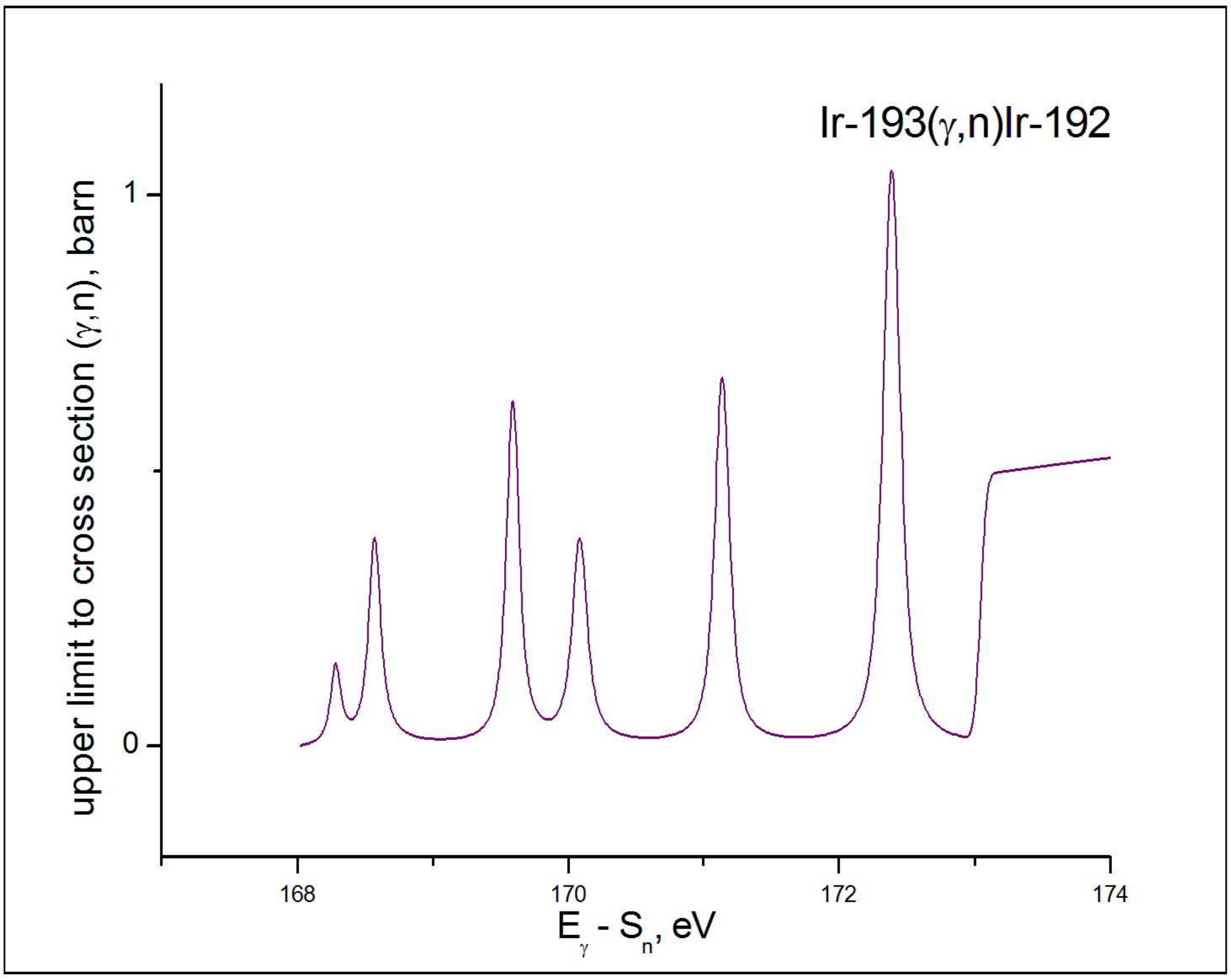

Fig. 7. (γ,n) resonances and upper limits to the integrated cross section for the reaction $^{193}$Ir(γ,n) $^{192}$Ir.

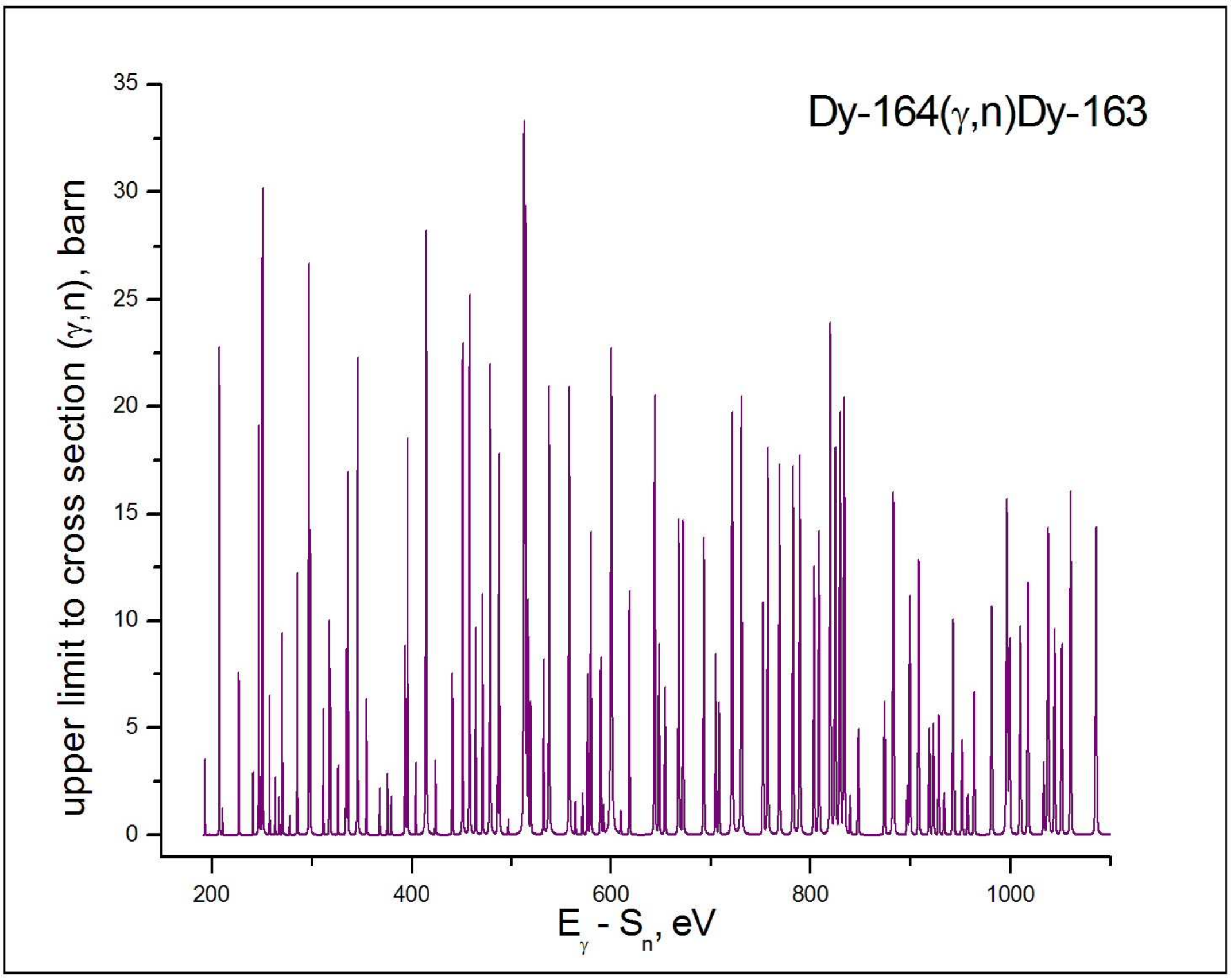

Fig. 8. (γ,n) resonances and upper limits to the integrated cross section for the reaction $^{164}Dy(\gamma,n)^{163}Dy$.

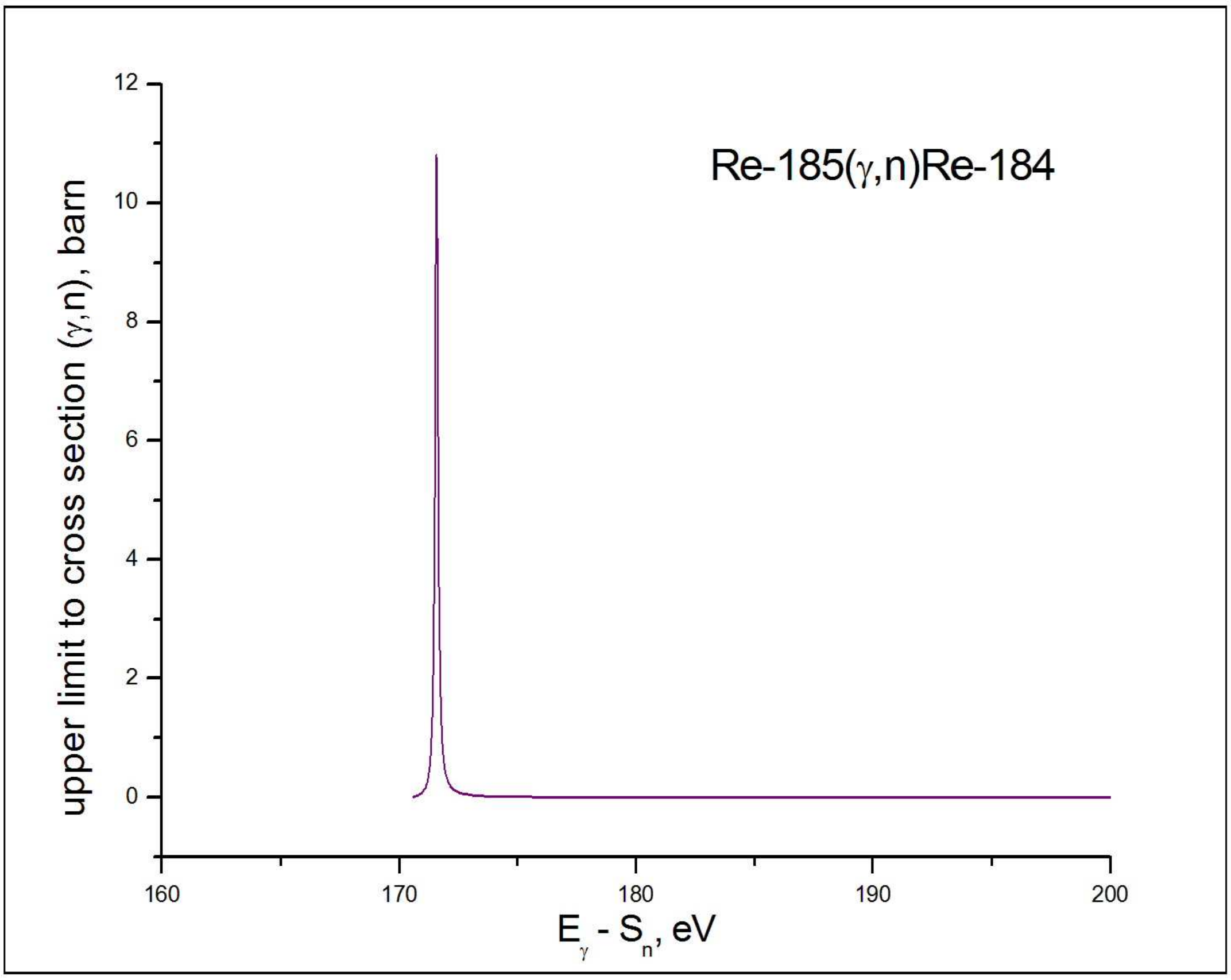

Fig. 9. (γ,n) resonances and upper limits to the integrated cross section for the reaction $^{185}Re(\gamma,n)\,^{184}Re$.

### 2.2 Unstable isotopes

The unstable nuclei for which we could find thermal-neutron resonances are listed in Table II up to energies of the emitted neutrons of 0.1 eV. Upper limits for the corresponding integrated cross sections, with the emission of neutrons into 4π, are also given in Table II.

| Reaction | Upper limit for the integrated cross section, b-eV | Position of resonance $E_\gamma - S_n$, eV | Width of resonance, eV | Height of resonance, b | Lower limit of the neutron energy $D_-$, eV | Upper limit of the neutron energy $D_+$, eV | Half-life of target nuclei |
|---|---|---|---|---|---|---|---|
| $^{242}Am(\gamma,n)^{241}Am$ | 0.014 | 68.3 | 0.056 | 1.786E-1 | 4.317E-4 | 1.178 | 16.02 h |
| $^{243}Am(\gamma,n)^{242}Am$ | 7.370E-4 | 89.9 | 0.589 | 1.090E-3 | 4.887E-4 | 1.539 | 7370 y |
| $^{250}Bk(\gamma,n)^{249}Bk$ | 0.025 | 53.2 | 0.042 | 4.060E-1 | 4.915E-4 | 0.814 | 3.212 h |
| $^{192}Ir(\gamma,n)^{191}Ir$ | 0.025 | 108.1 | 0.086 | 1.946E-1 | 0.00272 | 2.417 | 73.827d |
| $^{152}Eu(\gamma,n)^{151}Eu$ | 0.021 | 141.6 | 0.110 | 1.241E-1 | 0.00306 | 3.948 | 13.537 y |
| $^{155}Eu(\gamma,n)^{154}Eu$ | 0.026 | 231.6 | 0.155 | 1.172E-1 | 0.00373 | 5.699 | 4.753 y |
| $^{233}Pa(\gamma,n)^{232}Pa$ | 0.002 | 98.5 | 0.369 | 5.990E-3 | 0.00449 | 1.529 | 26.975 d |
| $^{232}Pa(\gamma,n)^{231}Pa$ | 0.007 | 71.6 | 0.057 | 8.812E-2 | 0.00548 | 1.408 | 1.32 d |
| $^{240}Pu(\gamma,n)^{239}Pu$ | 0.010 | 95.8 | 0.100 | 5.723E-2 | 0.00778 | 1.388 | 6561 y |
| $^{230}Th(\gamma,n)^{229}Th$ | 0.060 | 108.3 | 0.060 | 7.029E-1 | 0.00802 | 2.143 | 75400 y |
| $^{242}Pu(\gamma,n)^{241}Pu$ | 0.010 | 88.5 | 0.114 | 5.689E-2 | 0.00885 | 1.252 | 375000 y |
| $^{236}U(\gamma,n)^{235}U$ | 0.002 | 97.7 | 0.240 | 4.610E-3 | 0.01164 | 1.398 | 2.342E7 y |
| $^{244}Am(\gamma,n)^{243}Am$ | 1.639E-4 | 63.8 | 0.095 | 1.100E-3 | 0.01745 | 1.334 | 10.1 h |
| $^{254}Cf(\gamma,n)^{253}Cf$ | 0.035 | 77.3 | 0.129 | 1.719E-1 | 0.01823 | 1.536 | 60.5 d |
| $^{232}Pa(\gamma,n)^{231}Pa$ | 0.002 | 71.7 | 0.060 | 2.572E-2 | 0.02044 | 1.577 | 1.32 d |
| $^{237}Np(\gamma,n)^{236}Np$ | 4.203E-4 | 98.2 | 0.271 | 1.560E-3 | 0.02151 | 1.310 | 2.144E6 y |
| $^{114}In(\gamma,n)^{113}In$ | 0.059 | 251.1 | 0.115 | 3.670E-1 | 0.02442 | 7.926 | 71.9s |
| $^{233}Pa(\gamma,n)^{232}Pa$ | 0.001 | 98.8 | 0.531 | 3.400E-3 | 0.02532 | 2.140 | 26.975 d |
| $^{199}Au(\gamma,n)^{198}Au$ | 1.113 | 155.7 | 0.202 | 3.597E0 | 0.02565 | 2.604 | 3.139 d |
| $^{238}Np(\gamma,n)^{237}Np$ | 0.006 | 68.4 | 0.052 | 8.015E-2 | 0.02597 | 1.522 | 2.117 d |
| $^{116}In(\gamma,n)^{115}In$ | 0.870 | 214.6 | 0.108 | 5.823E0 | 0.02717 | 6.549 | 14.1 s |
| $^{156}Eu(\gamma,n)^{155}Eu$ | 1.826 | 139.0 | 0.101 | 1.220E1 | 0.03027 | 2.951 | 15.19 d |
| $^{169}Yb(\gamma,n)^{168}Yb$ | 0.043 | 150.4 | 0.076 | 3.721E-1 | 0.03128 | 2.940 | 32.018 d |
| $^{237}Np(\gamma,n)^{236}Np$ | 0.004 | 98.7 | 0.141 | 1.634E-2 | 0.03578 | 2.191 | 2.144E6 y |
| $^{250}Cf(\gamma,n)^{249}Cf$ | 0.121 | 94.9 | 0.162 | 4.775E-1 | 0.04735 | 2.103 | 13.08 y |
| $^{147}Pm(\gamma,n)^{146}Pm$ | 1.775 | 215.3 | 0.148 | 7.950E0 | 0.04798 | 4.871 | 2.6234 y |
| $^{242}Am(\gamma,n)^{241}Am$ | 0.023 | 68.6 | 0.060 | 2.724E-1 | 0.04980 | 1.658 | 16.02 h |
| $^{227}Ra(\gamma,n)^{226}Ra$ | 0.002 | 49.7 | 0.045 | 3.234E-2 | 0.06853 | 1.432 | 42.2 m |
| $^{252}Cf(\gamma,n)^{251}Cf$ | 0.148 | 81.8 | 0.134 | 3.736E-1 | 0.07371 | 1.990 | 2.645 y |
| $^{152}Eu(\gamma,n)^{151}Eu$ | 0.058 | 141.0 | 0.101 | 4.454E-1 | 0.08511 | 2.689 | 13.537 y |
| $^{254}Es(\gamma,n)^{253}Es$ | 0.079 | 55.4 | 0.038 | 1.551E0 | 0.08733 | 1.508 | 275.7 d |
| $^{232}Pa(\gamma,n)^{231}Pa$ | 0.001 | 72.0 | 0.062 | 1.254E-2 | 0.09127 | 2.002 | 1.32 d |
| $^{155}Eu(\gamma,n)^{154}Eu$ | 0.004 | 231.1 | 0.164 | 1.792E-2 | 0.09151 | 4.605 | 4.753y |
| $^{177}Lu(\gamma,n)^{176}Lu$ | 0.074 | 153.3 | 0.089 | 6.036E-1 | 0.09840 | 4.725 | 6.647 d |

Table II. Unstable-isotope (γ,n) resonances and upper limits for the integrated cross sections for the (γ,n) reaction, with the emission of the neutrons into 4π. $D_-$ represents the energy of the neutron emitted backwards with respect to the direction of incidence, and $D_+$ the energy of the neutron emitted along the direction of incidence of the γ-ray photon.

The upper limits for the integrated cross section correspond to the emission of the neutrons in all directions. The integrated cross sections for the generation of thermal neutrons are smaller than these limits by a factor of about 1/36, and can also be smaller than the listed limits because of the branching ratios.

## 3. Generation of thermal neutrons via (γ,n) reactions

The upper limit for the number of thermal neutrons generated per second via the (γ,n) reaction for a target having a thickness of $3x10^{-2}$ cm and for 400 incident gamma-ray photons/eV/s, a spectral intensity given in ref. [2], is of the order of

($3x10^{22}$ nuclei/$cm^3$) x (2 barn-eV) x ($3x10^{-2}$ cm) x (400 photons/eV/s) = 0.72 thermal neutrons/second,

in all directions. The thermal neutrons represent a small fraction of the total number of generated photoneutrons.

## 4. Conclusions

In this work we have determined the positions of the (γ,n) resonances and upper limits for the integrated cros sections for the (γ,n) reactions with the emission of neutrons in 4π, using data for the inverse process (n,γ).

We have identified the cases for which a (γ,n) resonance has a position such that the neutron emitted backwards with respect to the direction of incidence of the gamma-ray photon may have zero energy, and we have estimated the rate of generation of thermal neutrons via this mechanism.

Among the reactions producing thermal neutrons via (γ,n) reaction we mention $^{185}$Re(γ,n)$^{184}$Re with an upper limit of the integrated cross section of 2.4 b-eV, and $^{178}$Hf(γ,n)$^{177}$Hf with an upper limit of the integrated cross section of 0.9 b-eV, in all directions.